\documentstyle[12pt]{amsart}
\def\bR{{\Bbb R}}
\def\bZ{{\Bbb Z}}
\def\bC{{\Bbb C}}
\def\bP{{\Bbb P}}
\def\bH{{\Bbb H}}
\def\bQ{{\Bbb Q}}

\def\ci{{\sqrt{-1}}}
\def\Gikj{{\Gamma^{i}_{kj}}}
\def\Gjki{{\Gamma^{j}_{ki}}}

\def\Glki{{\Gamma^{l}_{ki}}}
\def\Gjkm{{\Gamma^{j}_{km}}}
\def\Gjmk{{\Gamma^{j}_{mk}}}
\def\Glmi{{\Gamma^{l}_{mi}}}
\def\Glit{{\Gamma^{l}_{it}}}
\def\Di{{\frac{\partial}{\partial x^i}}}
\def\Dk{{\frac{\partial}{\partial x^k}}}
\def\Dm{{\frac{\partial}{\partial x^m}}}
\def\Dl{{\frac{\partial}{\partial x^l}}}
\def\Dj{{\frac{\partial}{\partial x^j}}}
\def\aleph{{\mathcal S}}
\def\oY{{\overline Y}}\def\cL{{\mathcal L}}
\newcommand{\PR}{{\noindent\rm\bf Proof.}}
\newcommand{\EndP}{\hfill\square\\}
\newtheorem{thm}{Theorem}[section]
\newtheorem{definition}[thm]{Definition}
\newtheorem{lemma}[thm]{Lemma}
\newtheorem{proposition}[thm]{Proposition} 
\newtheorem{corollary}[thm]{Corollary}
\newtheorem{remark}[thm]{Remark}
\newtheorem{example}[thm]{Example}
\title{Some remarks on special K\"ahler geometry}
\author[]{Claudio Bartocci}\address{Universit\`a degli Studi di Genova, Dipartimento di Matematica, Via Dodecaneso 35, 16146 Genova, Italy}\email{bartocci@@dima.unige.it}
\author[]{Igor Mencattini}\address{Universit\"at Augsburg, Instit\"ut f\"ur Mathematik, Augsburg 86135, Germany}\email{Igor.Mencattini@@math.uni-augsburg.de}

\begin{document}\maketitle

\begin{abstract} Given a special K\"ahler manifold $M$, we give a new, direct proof of the relationship between the quaternionic structure on $T^\ast M$  and the variation of Hodge structures on $T^{\bC}M$.
\end{abstract}

\section{Introduction}

Our aim, in this paper, is to make explicit some connections between special geometry, quaternionic geometry and variations of Hodge structures, which remain somehow hidden in the physical and mathematical literature on the subject.\\
As a motivation from a physical point of view,
affine special K\"ahler geometry can be thought as the geometry of the quantum moduli spaces for $N=2$ supersymmetric Yang-Mills (SYM) theory in four-dimensional space ~\cite{Do}, which is the setting for Montonen-Olive duality ~\cite{MO} and for the work of Seiberg-Witten ~\cite{SW1, SW2}. An important property of these theories is that their low-energy behavior is encoded in a finite number of physical parameters, which become the coordinates $(x_1,\ldots,x_{2n})$ for the quantum moduli space of the theory. As a consequence of supersymmetry, all meaninful physical quantities are (locally defined) holomorphic functions of the coordinates. Among those, particular relevant are the electric charge $z=(z_1,\ldots,z_n)$, whose components can be used as local holomorphic coordinates, and the effective lagrangian ${\mathcal L}_{eff}$. Supersymmetry allows to write ${\mathcal L}_{eff}$ 
in terms of a single holomorphic function ${\mathcal F}={\mathcal F}(z_1,\ldots,z_n)$ called the prepotential. All other physical quantities can be deduced from the functions $\mathcal F$ and $(z_1,\ldots,z_n)$. In particular, the magnetic charge of the theory (dual to the electric charge $z$) can be derived as:
$$w=\frac{\partial{\mathcal F}}{\partial z}$$
and the Montonen-Olive duality can be expressed by saying that:
$${\mathcal L}_{eff}(z)={\tilde {\mathcal L}}_{eff}(w)$$
where ${\tilde {\mathcal L}}_{eff}(w)$ is a new (effective) lagrangian in the (holomorphic) coordinates $w=(w_1,\ldots,w_n)$. The jacobian of the change of coordinates $\tau(z)=\frac{\partial w}{\partial z}=\frac{\partial^2{\mathcal F}}{\partial z^2}$ plays the role of (complex) coupling constant of the the theory. This is related to real coupling constant $g$ of the theory and to the phase angle $\theta$ by the formula, ~\cite{SW1}:
$$\tau=\frac{\theta}{2\pi}+\frac{4\pi\ci}{g^2}$$

The symmetries of the theory are given by the transformations $\tilde\tau=\frac{\partial (-z)}{\partial w}=-\frac{1}{\tau(z)}$ and the action of a finite index subgroup of $Sp(2n,\bZ)$ ~\cite{SW1, SW2}.\\
In ~\cite{SW1, SW2} the authors interpret the presence of all charges (electric and magnetic) in terms of a unimodular lattice smoothly changing with the point $(x_1,\ldots,x_{2n})$ on the moduli space of the theory. This lattice is interpreted as the homology bundle of a family of (hyper)elliptic curves, whose relative jacobian carries a natural (holomorphic) symplectic form, whose fibers are lagrangian. From this description, it stems the relevance of the theory of integrable systems for the SYM-theory in four dimensions ~\cite{DW},\cite{Fr}.\\
All the relevant physical quantities just introduced, have a nice geometrical description in terms of special geometry, whose main features we review in Section 2.

The close connection between those geometrical structures and variations of Hodge structures --- which we deal with in Section 3 --- was already pointed out by Seiberg and Witten in their seminal papers  ~\cite{SW1, SW2} (see also the nice review ~\cite{Do} by Donagi). Further developments are due to  Freed ~\cite{Fr} and to Hertling ~\cite{He}, where an elegant account of the far-reaching generalization of the special geometry known as $tt^{\ast}$-geometry is presented.

A special K\"ahler structure on $M$ induces a quaternionic (actually, hyperk\"ahler strucure) on $T^\ast M$. This fact, already  already present in the physical literature (see e.g. ~\cite{CFG}),  have been rigorously proved by Freed ~\cite{Fr}. In theorem (\ref{mainth}) we 
prove that the quaternionic structure on $T^\ast M$ can be directly associated to the variation of Hodge structures on $T_{\bC}M$ induced by the special K\"ahler geometry on $M$. To prove this result we rely on some constructions given by Simpson in ~\cite{Si1} and ~\cite{Si2}.

For different viewpoints on special geometry we refer to the papers ~\cite{ACD} and ~\cite{C}.

\section{Special K\"ahler manifolds}

\subsection{Affine special geometry}

In this section we review the basic facts about special K\"ahler geometry. Our main references are Cortes ~\cite{C}, Freed ~\cite{Fr}, Hitchin ~\cite{Hi} and Salamon ~\cite{Sa} 

\begin{definition}\label{basicdef} $(M,I,\nabla)$ is a special complex manifold if $(M,I)$ is a complex manifold and $\nabla$ is a flat, torsion free linear connection with the property $d_{\nabla}I=0$.\\
$(M,I,\nabla,\omega)$ is a special K\"ahler manifold if $(M,I,\omega)$ is a K\"ahler manifold, $(M,I,\nabla)$ is a special complex manifold and $\nabla\omega=0$
\end{definition}

\begin{remark}\rm
1) The linear connection $\nabla$ is defined on the real tangent bundle. Its $\bC$-linear extension to $T^{\bC} M$, the complexification of the tangent bundle, does not preserve the decomposition $T^{\bC} M=T^{1,0}\oplus T^{0,1}$.\\
2) Since $I\in\Omega^{1}(TM)$, $d_{\nabla}I\in\Omega^{2}(TM)$. In particular, $d_{\nabla}I=0$ if and only if $d_{\nabla}I(X,Y)=0$ for every pair of vector fields $(X,Y)$.
\end{remark}

Let $M$ be a manifold, $\nabla$ a torsion free connection on $TM$ and $I\in \Omega^0(End(TM))$.

\begin{lemma}\label{lemmino}
$d_{\nabla}I=0$ if and only if $(\nabla_XI)Y=(\nabla_YI)X$ for every $X,Y$ vector fields on $M$.
\end{lemma}
$\PR$ 
In local coordinates $(x_1,\cdots,x_n)$, we have: $\nabla_{\Dk}{\Dj}={\Gikj}{\Di}$,$\nabla_{\Dk}dx^{j}=-{\Gjki}dx^i$ and $I=I^{i}_{j}dx^j\otimes{\Di}$. Then:

$$\big(\nabla_{\Dk}I\big){\Dm}=\frac{\partial I^{i}_{m}}{\partial x^{k}}{\Di}-I^{i}_{j}{\Gjkm}{\Di}+I^{i}_{m}{\Glki}{\Dl}$$ 
$$\big(\nabla_{\Dm}I\big){\Dk}=\frac{\partial I^{i}_{k}}{\partial x^{m}}{\Di}-I^{i}_{j}{\Gjmk}{\Di}+I^{i}_{k}{\Glmi}{\Dl}.$$

Moreover:
$${\tau}_{\nabla}=0\iff {\Gjkm}={\Gjmk}\:\:\:\forall j,k,m$$and:$$\big(\nabla_{\Dk}I\big){\Dm}-\big(\nabla_{\Dm}I\big){\Dk}=\big(\frac{\partial I^{i}_{m}}{\partial x^{k}}-\frac{\partial I^{i}_{k}}{\partial x^{m}}\big){\Di}+\big(I^{i}_{m}{\Glki}-I^{i}_{k}{\Glmi}\big){\Dl}.$$

On the other hand:
$$d_{\nabla}I=d_{\nabla}(I^{i}_{j}dx^{j}\otimes{\Di})=\frac{\partial I^{i}_{j}}{\partial x^t}dx^{t}\wedge dx^{j}\otimes{\Di}-I^{i}_{j}{\Glit}dx^{j}\wedge dx^{t}\otimes{\Dl}$$ 
so that:
$$d_{\nabla}I\big({\Dk},{\Dm}\big)=\big(\frac{\partial I^{i}_{m}}{\partial x^{k}}-\frac{\partial I^{i}_{k}}{\partial x^{m}}\big){\Di}+\big(I^{i}_{m}{\Glki}-I^{i}_{k}{\Glmi}\big){\Dl}.$$
$\EndP$

\begin{lemma}\label{f}
Every special complex manifold $(M,I,\nabla)$ (special K\"{a}hler manifold) admits an affine structure whose transition functions take values in Gl$(2n,{\bR})$ (Sp$(2n,{\bR})$).
\end{lemma}
$\PR$ The linear connection on $TM$ defines a linear connection on $T^{\ast}M$. Since the connection is flat, for each point $x\in M$, we can find a basis $\alpha_1,\cdots,\alpha_{2n}$ of local sections of $T^{\ast}M$ defined over a suitable open neighborhood of $x$, such that $\nabla\alpha_i=0$. Since the connection is torsion free, the 1-forms $\alpha_i$ are closed and then locally exact. The local potentials for the $\alpha_i$'s can be used as local coordinates around the point $x$. Given another set of such coordinates, the transition functions will take values in Gl$(2n,{\bR})$. If $M$ is special K\"{a}hler, the condition $\nabla\omega=0$, implies that we can choose the local potential of the flat sections $(\alpha_1,\cdots,\alpha_{2n})$ as canonical coordinates for the symplectic form $\omega$. From this the statement follows. 
$\EndP$

\begin{lemma} 
Any set of flat coordinates (as the one defined in lemma ({\ref f})) on a special complex manifold defines a distinguished double set of holomorphic coordinates $(z_1,\dots,z_n)$ and $(w_1,\dots,w_n)$.\end{lemma}
$\PR$ 
We note that the flatness of the connection $\nabla$ together with the condition $d_{\nabla}I=0$ imply that there exists a (locally defined) vector field $X$ such that $\nabla(X)=I$. Given a local set of flat coordinates $(x_{1},\dots,x_{n},x_{n+1},\cdots,x_{2n})=(x_{1},\dots,x_{n},y_{1},\cdots,y_{n})$ we can write:
$$X=\sum_{k=1}^{n}\big(p_{k}\frac{\partial}{\partial x_k}+q_{k}\frac{\partial}{\partial y_k}\big)$$
for suitable functions (locally defined) $p_k,q_k$ $k=1,\cdots,n$. With respect to this choice of coordinates, we can write the complex structure as:
$$I=\sum_{k=1}^{n}\big(dp_{k}\otimes\frac{\partial}{\partial x_k}+dq_{k}\otimes\frac{\partial}{\partial y_k}\big)\,.$$ By using this representation of the complex structure we can define $z_j=x_j+{\ci}p_j$ and $w_j=y_j+{\ci}q_j$, $j=1,\cdots,n$. This ends the proof. $\EndP$

The coordinates $(z_1,\cdots,z_n)$ and $(w_1,\cdots,w_n)$ introduced in the previous lemma are related as follows:
$$dw_{r}=\sum_{s=1}^{n}{\tau}_{rs}dz_{s}\,,$$
where $\tau_{rs}=\tau_{rs}(z_1,\cdots,z_n)$ describes a holomorphic change of coordinates.

\begin{lemma}
The holomorphic 2-form $\Omega=\sum_{r=1}^{n}dw_r\wedge dz_{r}$ is identically equal to zero.
\end{lemma}
$\PR$
Since $I$ is a symplectic endomorphism of $TM$ we have:
$$\sum_{i=1}^{n}dx_i\wedge dy_{i}=\omega=I(\omega)=\sum_{i=1}^{n}I(dx_i)\wedge I(dy_{i})=\sum_{i=1}^{n}dp_i\wedge dq_{i}$$
From this, it follows that $Re(\Omega)=\sum_{i=1}^{n}dx_i\wedge dy_{i}-\sum_{i=1}^{n}dp_i\wedge dq_{i}=0$, which implies that $\Omega=-{\overline\Omega}$. The statement follows from this last equality.
$\EndP$

From the previous lemma we conclude that the one form $\theta=\sum_{r=1}^{n}w_{r}dz_{r}$ is closed so that there exists a (locally defined) holomorphic function ${\mathcal F}={\mathcal F}(z_1,\ldots,z_1)$, such that $\theta=d{\mathcal F}$. In particular we have:
$$w_r=\frac{\partial{\mathcal F}}{\partial {z}_r},\:\:\forall\:\: r=1,\ldots,n$$

\begin{definition}
The function $\mathcal F$ is called the prepotential of the special K\"ahler manifold $M$.
\end{definition}

\begin{example}\rm
Let us consider  the cotangent bundle  $T^\ast\bC^{n} = \bC^{2n}$ of $\bC^{n}$, with coordinates $(z_1,\ldots,z_n, w_1,\dots, w_n)$, endowed with its  canonical symplectic form $\Omega=\sum_{i=1}^{n}dz_{i}\wedge dw_i$.\\
Every simply connected special K\"ahler manifold can be realized as a lagrangian submanifold of $(T^\ast\bC^{n},\Omega)$.\\
The prepotential is the generating function of the holomorphic Lagrangian immersion.
\end{example}

The following theorem explains the significance of special manifolds in the context of quaternionic geometry:

\begin{thm}\label{sk} ~\cite{Fr}
If $M$ is a special K\"ahler manifold then its cotangent bundle is hyperk\"ahler.
\end{thm}
It is not difficult to write down the  quaternionic structure on the fibres $T_\alpha(T^\ast M)$. The flat connection $\nabla$ induces the decomposition
\begin{equation} T_\alpha(T^\ast M) \simeq T_{\pi(\alpha)} M \oplus
T^\ast_{\pi(\alpha)} M \simeq T_{\pi(\alpha)} M \oplus \overline{T_{\pi(\alpha)} M}\end{equation}
where $\pi\colon T^\ast M \to M$ is the canonical projection. We define
\begin{equation} \label{jstructure}
J(v , \bar{w}) = (-w,  \bar{v})
\end{equation}
It turns out that the complex structure $J$ is integrable and 
that $I\circ J = - J\circ I$ (see \cite{Fr} for details).

\section{Hodge structures and variations of Hodge structure} 

We begin by recalling some basic definitions (see e.g.~\cite{G-S}).\\

A real structure $\mathfrak r$ on a complex vector space $V$ is $\bR$-linear map anticommuting with the multiplication by $\ci$.

\begin{example}\rm
If $V=V_{\bR}\otimes_{\bR}{\bC}$ is the complexification of  a real vector space $V_{\bR}$, then the complex conjugation is one example of real structure,
\end{example}

Let $n$ be the complex dimension of $V$ and $p\leq n$ a nonnegative integer.

\begin{definition} 
A Hodge structure of weight $p$ on $V$ is given by a direct sum decomposition $V=\bigoplus_{r+s=p}V^{r,s}$  such that $V^{r,s}= {\mathfrak r} (V^{s,r})$.
\end{definition}


\begin{lemma}
A Hodge's structure of weight $p$ on $V$ is equivalent to a decreasing filtration, called Hodge filtration:
$$0\subset\cdots\subset{\mathcal F}^{k+1}\subset{\mathcal F}^{k}\subset{\mathcal F}^{k-1}\subset\cdots\subset V$$
where ${\mathcal F}^s=\oplus_{i\geq k}V^{i,p-i}$.
\end{lemma}
$\PR$(sketch) The Hodge filtration determines the Hodge structure:
$$V^{k,l}={\mathcal F}^k\bigcap{\overline{\mathcal F}}^l.$$
On the other hand, a decreasing filtration on $V$ is a Hodge filtration for some Hodge structure of weight $p$ on $V$, if and only if there is an isomorphism:
$$ {\mathcal F}^k\oplus{\overline {\mathcal F}}^{p-k+1}\longrightarrow V$$
for each $k$. $\EndP$

\begin{example}\rm 
Let $X$ be a complex K\"{a}hler manifold and let $H^{\bullet}(X,\bC)$ be its complex cohomology ring. Then for each $n$, $H^n(X,\bC)$ carries a Hodge structure of weight $n$: $H^n(X,\bC)=\bigoplus_{k+l=n}H^{k,l}(X)$, $H^{k,l}(X)={\overline H^{l,k}(X)}$ where the $H^{k,l}(X)$'s are the Dolbeault cohomology groups of $X$.
\end{example}


Let $V=\bigoplus_{k+l=p}V^{k,l}$ be a Hodge structure of weight $p$ and $Q:V\otimes V\longrightarrow\bC$ a bilinear, nondegenerate form, rational on $V_{\bZ}$. The form $Q$ is assumed to be symmetric if $p$ is even and antisymmetric if $p$ is odd.

\begin{definition}
The Hodge structure on $V$ is said to be polarized by $Q$ if  $Q(V^{k,l},V^{r,s})=0$ unless $k=r$ and $l=s$ and $({\ci})^{k-l}Q(x,{\overline x})>0$ for every $x\in V^{k,l}$, $x\neq 0$.
\end{definition}

Let us now introduce the notion(s) of variation of (polarized) Hodge structures. 

Let $M$ be a complex manifold.

\begin{definition}\label{VHPS}
By a variation of polarized Hodge structures (VPHS) with base $M$ and weight $p$, we shall mean the following set of data:\\
1) a complex vector bundle ${\mathcal H}\longrightarrow M$
endowed with a flat connection $\nabla:{\mathcal O}({\mathcal H})\longrightarrow \Omega^1\otimes{\mathcal O}({\mathcal H})$, and
a (real) flat vector subbundle ${\mathcal H}_{\bR}\subset {\mathcal H}$ together with a flat bundle ${\mathcal  H}_{\bZ}\subset{\mathcal H}_{\bR}$ such that $({\mathcal H}_{\bZ})_m\bigcap({\mathcal H}_{\bR})_m$ is a lattice for each $m\in M$;\\
2) a flat bilinear form $Q:{\mathcal O}({\mathcal H})\times{\mathcal O}({\mathcal H})\longrightarrow{\bC}$ such that $Q(x,y)=(-1)^pQ(y,x)$ and $Q\vert_{{\mathcal O}({\mathcal H}_{\bZ})\times{\mathcal O}({\mathcal H}_{\bZ})}\in{\bQ}$;\\
3) a decreasing filtration $0\subset\cdots\subset{\mathcal V}^{k+1}\subset{\mathcal V}^{k}\subset{\mathcal V}^{k-1}\subset\cdots\subset{\mathcal V}^0={\mathcal H}$ with the property: $\nabla({\mathcal O}({\mathcal V}^k))\subset{\mathcal O}({\mathcal V}^{k-1})\otimes T^{\ast}M$.
\end{definition}

The data in the definition above need to satisfy the condition that for each $m\in M$, the filtration ${\mathcal V}^{\bullet}\vert_{m}$ defines a Hodge structure of weight $p$ on the fiber $H_m$. Such a Hodge structure is polarized by the restriction to ${\mathcal H}_m$ of the bilinear form $Q$.

\begin{remark}\rm
The bilinear form $Q$ defines a (in general indefinite) hermitian metric on the bundles ${\mathcal V}^k$ of the filtration $\mathcal V^{\bullet}$.
\end{remark}

We begin discussing the relation between special geometry and variation of Hodge structures proving the following theorem, Hertling ~\cite{He}:

\begin{thm}
There is a one to one correspondence between VHS of weight $1$ on $T^{\bC} M$ and special complex structures on $M$. Such a correspondence extends naturally to a correspondence between VPHS    on $T^{\bC} M$ and special K\"ahler structures on $M$.
\end{thm}
$\PR$ 
The flat and torsion free connection $\nabla$ defined on $TM$, extends to a torsion free and flat connection $\nabla$ to $T^{\bC}M$, which defines a holomorphic structure on $T^{\bC}M$ (note that, since $\nabla^{2}=0$, $(\nabla^{0,1})^{2}=0$). To prove the theorem, it suffices to show that $T^{1,0}$ is a holomorphic subbundle of $T^{\bC}M$. In fact, this will provide the (decreasing) flag of holomorphic bundles: ${\cal F}^2=\{0\}\subset{\cal F}^1= T^{1,0}\subset{\cal F}^0=T^{\bC}M$.\\
We are left to show that $\nabla^{0,1}(X)=0$ for each $X\in\Gamma(T^{1,0})$, which is equivalent to show that $\nabla_{\overline Y}(X)=0$ for each $X,Y\in\Gamma(T^{1,0})$.\\
By lemma (\ref{lemmino}), since $d_{\nabla}I=0$, we have that $(\nabla_{\xi}I)\eta=(\nabla_{\eta}I)\xi$ for every $\xi,\eta\in\Gamma(T^{\bC}M)$. In particular, $(\nabla_{X}I){\overline Y}=(\nabla_{\overline Y}I)X$ for each $X,Y\in\Gamma(T^{1,0})$. Suppose that $X$ is holomorphic vector field. From the integrability of $I$, it follows that ${\mathcal L}_{X}I=0$. This implies the following:
$$0={\mathcal L}_{X}(I{\oY})-I({\cL}_{X}({\oY})=[X,I{\oY}]-I[X,{\oY}]=\nabla_{X}(I{\oY})-\nabla_{I{\oY}}(X)-I(\nabla_{X}{\oY})+I(\nabla_{\oY}X)=$$ $$(\nabla_{X}I){\oY}-\nabla_{I{\oY}}(X)+\nabla_{\oY}(IX)-(\nabla_{\oY}I)X=(\nabla_XI){\oY}-(\nabla_{\oY}I)X+2{\ci}(\nabla_{\oY}X)$$
$\EndP$

We will now write a set of equations that characterize (at a large extent) a special K\"{a}hler manifold, Freed ~\cite{Fr}.\\
Suppose that $(M,I,\nabla)$ is a special K\"{a}hler manifold. In this case a (pseudo) riemannian metric and its Levi-Civita connection $D$ are defined on $M$. Such a connection preserves the decomposition in the $(1,0)$, $(0,1)$ components of the bundle $T^{\bC}M$ (actually, the complex structure $I$ is constant with respect to the parallel transport defined via $D$). We can define the one form: $A_{\bR}=\nabla-D$. Taking the $\bC$-linear extensions of the connections $D$ and $\nabla$ (which will be still denoted with the same letters), we get $A, {\overline A}\in\Omega_{\bC}^1(End(T^{\bC}M))$ such that $A_{\bR}=A+{\overline A}$ (in fact, let $V$ be a real vector space and $A\in End_{\bR}(V)$. Extend $A_{\bR}$ to a complex endomorphism of $ V_{\bC}=V\otimes_{\bR}{\bC}$ as follows $A(v\otimes\alpha)=A(v)\otimes\alpha$ and ${\overline A}(v\otimes\alpha)=A(v)\otimes{\overline\alpha}$. Then $A_{\bR}=\frac{A+{\overline A}}{2}$).\\
Going back to our case, we have the following theorem:

\begin{thm}\label{t1}
$$A\in\Omega^{1,0}(Hom(T^{1,0},T^{0,1}))$$ 
$${\overline A}\in\Omega^{0,1}(Hom(T^{0,1},T^{1,0})).$$
\end{thm}
$\PR$ 
Given $X,Y\in\Gamma(T^{1,0})$, we have:
$$(\nabla_{\overline X}I)Y={\ci}(\nabla_{\overline X}Y)-I(\nabla_{\overline X}Y)$$
$$(\nabla_{Y}I){\overline X}=-{\ci}(\nabla_{Y}{\overline X})-I(\nabla_{Y}{\overline X})$$
and:
$$0=(D_{\overline X}I)Y={\ci}(D_{\overline X}Y)-I(D_{\overline X}Y)$$

$$0=(D_{Y}I){\overline X}=-{\ci}(D_{Y}{\overline X})-I(D_{Y}{\overline X})$$

From the first two equalities we get:

$${\ci}(\nabla_{\overline X}Y+\nabla_{Y}{\overline X})=I([{\overline X},Y])$$ 
and from the second ones:
$${\ci}(D_{\overline X}Y+D_{Y}{\overline X})=I([{\overline X},Y])$$ 
which imply:
$$2{\ci}\nabla_{\overline X}Y={\ci}[{\overline X},Y]+I([{\overline X},Y])$$
and
$$2{\ci}D_{\overline X}Y={\ci}[{\overline X},Y]+I([{\overline X},Y])\,.$$
From these equalities it follows that $\nabla_{\overline X}Y=D_{\overline X}Y$, i.e:
$${\overline A}_{\overline X}(Y)=0$$
Similarly, $\nabla_{X}{\overline Y}-D_{X}{\overline Y}=0$, i.e:
$$A_{X}({\overline Y})=0$$
These prove that: 
$$A\in\Omega^{1,0}(Hom(T^{1,0},T^{\bC}M))$$
and
$$A\in\Omega^{0,1}(Hom(T^{0,1},T^{\bC}M))$$
Since both $\nabla$ and $D$ preserve the symplectic form $\omega$, $A\in {\mathfrak sp}(2n,\bR)$. In particular:
$$\omega(A_{X}Y,{\overline Z})+\omega(Y,A_{X}{\overline Z})=0$$
for each $X,Y,Z\in\Gamma(T^{1,0})$. This implies that: $\omega(Y,A_{X}{\overline Z})=0$ so that $\omega(A_{X}Y,{\overline Z})=0$, i.e.~ $A_{X}(Y)\in\Gamma(T^{0,1})$ and $A\in\Omega^{1,0}(Hom(T^{1,0},T^{0,1}))$.
$\EndP$

Since the connection $\nabla=D+A_{\bR}$ is flat, we have:$$0=\nabla^2=(D+A_{\bR})^2=D^2+D(A_{\bR})+A_{\bR}(D)+A_{\bR}\wedge A_{\bR}=D^2+d_{D}(A_{\bR})+A_{\bR}\wedge A_{\bR}$$Since $A_{\bR}=A\wedge{\overline A}$,
\begin{equation}
R_{D}+A\wedge A+{\overline A}\wedge A+{\overline A}\wedge{\overline A}+A\wedge{\overline A}+d_{D}(A_{\bR})=0\label{e1}
\end{equation}
where $R_D$ is the curvature od $D$. Let us now consider  the decomposition of the connection $D$ in its $(1,0)$ and $(0,1)$ parts: $d_D={\partial}_D+{\overline {\partial}}_{D}$. We notice that ${\overline{\partial}}_D={\overline{\partial}}$, because $D$ is the Chern connection of the K\"{a}hler form $\omega$. The decomposition in $(2,0)$, $(0,2)$ and $(1,1)$-type components reads as follows:
\begin{equation}
\partial_{D}(A)+A\wedge A=0\label{e2}
\end{equation}
\begin{equation}
{\overline{\partial}}({\overline A})+{\overline A}\wedge{\overline A}=0\label{e3}
\end{equation}
\begin{equation}
R_{D}+A\wedge{\overline A}+{\overline A}\wedge A+{\overline{\partial}}(A)+\partial_{D}({\overline A})=0\label{e4}
\end{equation}

\begin{proposition}\label{p1}
Given $A_{\bR}=\nabla-D= A+ {\overline A}$ as in theorem (\ref{t1}), one has:
\begin{equation}\partial_{D}(A)=0,\label{e5}
\end{equation}
\begin{equation}{\overline\partial}(\overline A)=0\label{e6}
\end{equation}
\begin{equation} R_{D}+{\overline{\partial}}(A)+\partial_{D}({\overline A})+{\overline A}\wedge A+A\wedge{\overline  A}=0\label{e7}
\end{equation}
\end{proposition}
$\PR$ Equations (\ref{e6}) and (\ref{e7}) follow from equations (\ref{e2}) and (\ref{e3}) together with the $A\wedge A=0={\overline A}\wedge{\overline A}$, which are consequences of theorem (\ref{t1}). $\EndP$

\begin{lemma}
If $A_{\bR}$ is as in proposition (\ref{p1}), then $A$ is holomorphic and the following equations hold:
\begin{equation}\partial_{D}({\overline A})=0\label{e8}\end{equation}
\begin{equation}R_D+A\wedge{\overline A}+{\overline A}\wedge A=0\label{e9}
\end{equation}
\end{lemma}

$\PR$
From the previous proposition:
$$R_{D}+{\overline{\partial}}(A)+\partial_{D}({\overline A})+{\overline A}\wedge A+A\wedge{\overline A}=0$$
Observe that:
$$R_{D}+{\overline A}\wedge A+A\wedge{\overline A}=0$$
and
$${\overline\partial}(A)+\partial_{D}({\overline A})=0$$
separately, since $R_{D}+{\overline A}\wedge A+A\wedge{\overline A}$ is ${\bC}$-linear while ${\overline\partial}(A)+\partial_{D}({\overline A})$ is $\bC$-antilinear. On the other hand, ${\overline\partial}(A)$ takes values in $\Gamma(T^{0,1})$ while $\partial_{D}({\overline A})$ take values in $\Gamma(T^{1,0})$, so that ${\overline\partial}(A)=0$ and $\partial_{D}({\overline  A})=0$ as declared in the statement. 
$\EndP$

Let $E$ be a holomorphic vector bundle on a complex manifold $M$ and let
$$\phi:\Omega^0(E)\longrightarrow\Omega^{1,0}(E)$$ 
a $C^{\infty}(M)$-linear map.

\begin{definition}
The map $\phi$ is called a Higgs field if $\phi$ is holomorphic and if it satisfies \\
$\phi\wedge\phi=0$.
\end{definition}

The following result is now clear:

\begin{proposition}
The map $A$ of theorem (\ref{t1}) is a Higgs field defined on the holomorphic vector bundle $T^{\bC}M$.
\end{proposition}

\begin{remark}\rm 
For an explicit description of the curvature, the K\"ahler potential and the Higgs field in terms of the prepotential we refer the reader to
Freed ~\cite{Fr}, Hitchin ~\cite{Hi}.
\end{remark}

\subsection{Rees-Simpson bundles}

 In this section we review a construction due to Simpson ~\cite{Si1, Si2}, which defines a correspondence between vector spaces with a complete (decreasing) filtration and certain vector bundles on the Riemann sphere\\
Let $V$ be a finite dimensional vector space with a complete decreasing filtration, i.e:
$$V={\mathcal F}^{0}V\supseteq {\mathcal F}^{1}V\supseteq\cdots\supseteq {\mathcal F}^{n-1}V=\{0\}$$

Let  $\bC$ the complex plane, $\bC^{\ast}$ the multiplicative group of complex numbers and ${\cal O}_{\bC^{\ast}}$ the sheaf of regular functions on $\bC^{\ast}$. 
If $i: \bC^{\ast}\longrightarrow \bC$ the standard inclusion map, we define $\aleph(V,{\mathcal F})$ as the the vector bundle on $\bC$ whose sheaf of sections is the subsheaf of the 
$i_{\ast}(V\otimes_{\bC}{\cal O}_{\bC^{\ast}})$, generated by the sections of the form $z^{-k}v_k$, where $v_k\in F^kV$ and $z$ is the standard coordinate on $\bC$.

Note that the bundle $\aleph(V,{\mathcal F})$ is endowed with a natural action of $\bC^{\ast}$, which is induced by the action of the group $\bC^{\ast}$ on $i_{\ast}(V\otimes_{\bC}{\cal O}_{\bC^{\ast}})$.
To see that this construction can be inverted, it is convenient to recall the following construction. Let $R$ be a $k$-algebra over a field $k$, and let be a field, and let $I\subset R$ be an ideal.
Given the decreasing filtration $R\supset I\supset I^2\supset I^3\cdots$, define the Rees algebra of $R$ with respect to $I$ as:
$${\mathcal R}(R,I)=\sum_{n=-\infty}^{+\infty}I^nz^{-n}$$
with the convention that $I^k=R$ for each $k\leq 0$. Note that ${\mathcal R}(R,I)=R[z,z^{-1}I]\subset R[z,z^{-1}]$ In particular, ${\mathcal R}(R,I)$ is a $k$ algebra that can be regarded as a 
$k[z]$ algebra.
Moreover:
$${\mathcal R}(R,I)/z{\mathcal R}(R,I)=gr_IR=R/I\oplus I/I^2\oplus\cdots$$
$${\mathcal R}(R,I)/(z-a){\mathcal R}(R,I)=R,\forall a\in k^{\ast}$$

In complete analogy, if $(V,{\mathcal F})$ is a vector space with a complete decreasing filtration, we define $B\subset {\bC}[z,z^{-1}]\otimes_{\bC}V$ as the $\bC[z]$-module generated by the elements of the form $z^{-k}v_k$, for $v_k\in {\mathcal F}^kV$. We recover $\aleph(V,{\mathcal F})$ (with a $\bC^{\ast}$ action) as the coherent sheaf on $\bC$ associated to the $\bC[z]$-module $B$. In terms of this second description, the fiber of $\aleph(V,{\mathcal F})$ over $0$ is identified with $Gr_{\mathcal F}V=\bigoplus{\mathcal F}^k/{\mathcal F}^{k+1}\simeq B/zB$.

Starting with $W$ a locally free coherent sheaf on $\bC$ endowed with a $\bC^{\ast}$-action, we get a vector space with a decreasing filtration, defining $V=W_1$, the fiber of $W$ over $1\in\bC$, and looking at the orders of zeros of the $\bC^{\ast}$-invariant sections.\\
In fact, if we define $B=W(\bC)$, the action of $\bC^{\ast}$ gives the trivialization of $W$ over $\bC^{\ast}\subset\bC$, which, if read in terms of $B$, is equivalent to the following isomorphism of $\bC[z]$-modules:
$$B\otimes_{\bC[z]}\bC[z,z^{-1}]\simeq V\otimes_{\bC}\bC[z,z^{-1}]$$
The quotient of both sides by the ideal $(z-1)$, gives the isomorphism:
$$V\simeq B/(z-1)B\simeq W_1$$
The inclusion of $\bC[z]$-modules: $B\subset V\otimes_{\bC}\bC[z,z^{-1}]$, defines a decreasing filtration in $V$, by saying that:
$${\mathcal F}^k=\{v\in V\vert z^{-k}v\in B\}$$
It easy to check that such a filtration is complete.\\
It is a routine exercise to show that the two correspondences are one the inverse of the other, so that we have proven:
\begin{proposition}
There is a one to one correspondence between the set of vector spaces with a complete decreasing filtration and the set of vector bundle over the affine line endowed with an action of the multiplicative group $\bC^{\ast}$ covering the standard action on the line. 
\end{proposition}

Now we suppose that on the vector space $V$ we have two complete decreasing filtrations: ${\mathcal F}$ and $\overline{\mathcal F}$, which are conjugate with respect to a real structure on $V$ (this is the case when $V$ carries a Hodge structure of weight $k$).
Let us consider the open covering of $\bP^1$ given by two copies of $\bC$ glue on $\bC^{\ast}$ by the map: $z\mapsto\frac{1}{z}$. The restrictions of the bundles $\aleph(V,{\mathcal F})$ and $\aleph(V,\overline{\mathcal F})$ to $\bC^{\ast}$ are trivial, and they are both isomorphic to $\bC^{\ast}\times V$. We can glue such restrictions on $\bC^{\ast}$ together using the map $(v,z)\mapsto (v,z^{-1})$. In this way, we get a vector bundle $\aleph(V,{\mathcal F},\overline{\mathcal F})$ on $\bP^1$, which is called the Rees-Simpson bundle associated to the triple $(V,{\mathcal F}, \overline{\mathcal F})$.




\begin{proposition} \cite{Si1}\label{Simpson}
The two filtrations $\mathcal F$ and $\overline{\mathcal F}$ define a decomposition of the vector space $V$ which is pure of weight $w$ if and only if the vector bundle $\aleph(V,{\mathcal F}, \overline{\mathcal F})$ is semistable of slope $w$.
\end{proposition}
$\PR$
Suppose first that the filtrations $\mathcal F$ and 
$\overline{\mathcal F}$ define a pure decomposition of weight $w$ of $V$, i.e $V=\bigoplus_{k+m=w}V^{k,m}$. In this case 
$(V,{\mathcal F},{\overline{\mathcal F}})$ can be written as the direct sum of triples $(V^{k,m},{\mathcal F}_{\vert V^{k,m}},\overline{\mathcal F}_{\vert V^{k,m}})$. In particular we have that on $V^{k,m}$, ${\mathcal F}^k=V^{k,m}$ and ${\mathcal F}^{k+1}=0$ and $\overline{\mathcal F}^m=V^{k,m}$ and $\overline{\mathcal F}^{m+1}=0$. It is easy to check that the Rees modules associated to $(V^{k,m},{\mathcal F}_{\vert V^{k,m}})$ and to $(V^{k,m},\overline{\mathcal F}_{\vert V^{k,m}})$ are generated, respectively, by the sections of $V^{k,m}\otimes_{\bC}{\mathcal O}_{\bC^{\ast}}$ with pole at zero of order at most $k$ and at infinity of order at most $m$. From this it follows:
$$\aleph(V^{k,m},{\mathcal F}_{\vert V^{k,m}},{\overline{\mathcal F}}_{\vert V^{k,m}})=V^{k,m}\otimes_{\bC}{\mathcal O}_{\bP^1}(k\cdot 0+m\cdot\infty)\simeq V^{k,m}\otimes{\mathcal O}_{\bP^1}(w)$$
since $k+m=w$. If we take the direct sum over the pairs $(k,m)$ we find that the bundle $\aleph(V,{\mathcal F},{\overline{\mathcal F}})$ is semistable of slope $w$.

To show the converse, let us remark that if $U$ is a vector space such that $\aleph(V,{\mathcal F},{\overline\mathcal F})\simeq U\otimes{\mathcal O}_{\bP^1}(w)$, then, by comparing the fibers at the point $1\in\bC$, it follows the canonical isomorphism $U\simeq V$. Let $W$ a vector space of dimension equal to one and let us suppose that on $W$ there are filtrations ${\mathcal F}_W$ and ${\overline{\mathcal F}}_W$ such that ${\mathcal F}^w_W=W$, ${\mathcal F}^{w+1}_W=0$ and  ${\overline{\mathcal F}}^0_W=W$, ${\overline{\mathcal F}}^1_W=0$. The line bundle
${\mathcal L}=\aleph(W,{\mathcal F}_W,{\overline\mathcal F}_W)$ is isomorphic to ${\mathcal O}_{\bP^1}(w)$ with an induced action of the group $\bC^{\ast}$ (note that different liftings of the action of the multiplicative group $\bC^{\ast}$ correspond to different choices of the pair $(k,m)$, with $k+m=w$). The bundle $Hom({\mathcal L},\aleph(V,{\mathcal F}_V,\overline{\mathcal F}_V))$ is isomorphic to $U\otimes_{\bC}{\mathcal O}_{\bP^1}$ and it carries a $\bC^{\ast}$-action. This induces an action of $\bC^{\ast}$ on the space of global sections $U$ compatible with the isomorphism
$\aleph(V,{\mathcal F}_V,\overline{\mathcal F}_V))\simeq U\otimes_{\bC}{\mathcal O}_{\bP^1}$. The eigenspace decomposition of $U=\bigoplus_m U^m$ defined by the action of the multiplicative group $\bC^{\ast}$ together with the isomorphism $U\simeq U$, define the decomposition $V=\bigoplus_{k+m=w}V^{k,m}$, with $V^{w-m,m}=U^m$. It easy to check that such a decomposition suffices to split the pair of filtrations $({\mathcal F},\overline{\mathcal F})$ to give the converse of the statement.
$\EndP$

\section{Quaternionic structures}

A quaternionic vector space is a left $\bH$-module. 
The real dimension of a quaternionic vector space is $4n$.

A real vector space $W$ endowed with a pair of complex structures $I$ and $J$ such that $I\circ J = - J\circ I$ has an induced structure of quaternionic vector space.

The following result is well known.
\begin{definition} 
Let $V$ be a complex vector space. A quaternionic structure on $V$ is a $\bR$-linear endomorphism $J$ such that, $J^2=-1$ and $J(\alpha v)={\overline\alpha}J(v)$, for each $v\in V$, $\alpha\in\bC$. \end{definition}

\begin{lemma}\label{corre}
Let $V$ be a complex vector space. There is a one to one correspondence between quaternionic structures on $V$ and Hodge structures of weight 1 on $V$.
\end{lemma}
$\PR$ 
Let us consider a pure Hodge structure of weight 1 on $V$, $V=V^{1,0}\oplus V^{0,1}$, $V^{0,1}={\mathfrak r}(V^{1,0})$,  ${\mathfrak r}$ being a real structure on $V$. For each $u=(v,w)\in V=V^{1,0}\oplus V^{0,1}$ define $J(u)=J(v,w)=(-{\mathfrak r}(w),{\mathfrak r}(v))$. Then $J\in End_{\bR}(V)$ and $J^2=-1$. If $I$ is the complex structure on $V$, $I(v,w)=(\ci v,-\ci w)$. Then $J(\ci(v,w))=J(\ci v,\ci w)=(\ci w,-\ci v)=-\ci(-w,v)=-\ci J(v,w)$. Let $J$ be a quaternionic structure on $V$. The complex structure of $V$ defines the decomposition $V=V^{1,0}\oplus V^{0,1}$ and ${\mathfrak r}(v,u)=(-J(w),J(v))$ defines the real structure. On the other hand, since $(V,I,J)$ is a quaternionic vector space, we can choose a basis $v_1,\cdots,v_{4n}$ such that $I(v_k)=\ci v_k$, $k=1,\cdots,4n$ and $J(v_{k})=-v_{2n+k}$, $J(v_{2n+k})=v_{k}$,$k=1,\cdots,2n$. Define $V'=span_{\mathbb{C}}\langle\zeta_1,\cdots,\zeta_n\rangle$ and $V''=span_{\mathbb{C}}\langle\zeta_{n+1},\cdots,\zeta_{2n}\rangle$ where $\zeta_k=v_k+\ci v_{k+n}$, $\zeta_{k+n}=v_{k+2n}+\ci v_{2k+2n}$, $k=1,\cdots,n$. Clearly, $V', V''$ are complex vectorspaces (w.r.t.~$I$) and $V=V'\oplus V''$. Let $c:V\longrightarrow V$ be the linear map defined as $c(\zeta_k)=v_k-\ci v_{k+n}$ and$c(\zeta_{k+n})=v_{k+2n}-\ci v_{2k+2n}$. Define ${\mathfrak r}:V\longrightarrow V$ via the following: ${\mathfrak r}(\zeta_k)=-c\circ J(\zeta_k)$ and ${\mathfrak r}(\zeta_{k+n})=c\circ J(\zeta_{k+n})$, $k=1,\cdots,n$. Then, $(V,V',V'')$ defines a Hodge structure of weight 1 whose real structure is given by $\mathfrak r$.\par\hfill $\EndP$

By proposition (\ref{Simpson}) and lemma (\ref{corre}) we conclude that:
\begin{corollary}
Given a vector space $V$, there is a one to one correspondence between Hodge structures of weight 1 on $V$,
quaternionic structures on $V$ and semistable bundle of slope $w$ on $\bP^1$.
\end{corollary}

\subsection{Twistors spaces}

Let $M$ be a quaternionic manifold. The set of almost complex structures defining the quaternionic structure of $M$ are in one to one correspondence with the points of the Riemann sphere $\bP^1$. The product manifold $Z(M)=M\times\bP^1$ carries a natural almost complex structure defined as follows. Let $I_0$ be the standard complex structure on $\bP^1$, and let $(m,\zeta)$ be a point on $Z(M)$. On the tangent space $T_{(m,\zeta)}Z(M)$ let us define the endomorphism ${\mathcal I}_{(m,\zeta)}=(I_{\zeta},I_0))$, where $I_{\zeta}$ is the complex structure on $M$ corresponding to the point $\zeta\in\bP^1$ (the action of ${\mathcal I}_{(m,\zeta)}$ on $T_{(m,\zeta)}Z(M)\simeq T_mM\times T_{\zeta}\bP^1$ is componentwise). It is clear that the assingment of such a linear map for each point in $Z(M)$ define an almost complex structure on $Z(M)$.
\begin{definition}~\cite{HKLR, Ka}
The manifold $Z(M)$ is called the twistor space of the quaternionic manifold $M$,.
\end{definition}

For the proof of the following theorem we refer to Kaledin ~\cite{Ka}:
\begin{thm}
Let $M$ be a quaternionic manifold and let $Z(M)$ be its twistor space. The following conditions are equivalent:\\
1) Two almost complex structures $I$ and $J$, with $J\neq I$ and $J\neq{\overline I}$ are integrable;\\
2) Every almost complex structure $I$ belonging to the quaternionic structure is integrable;\\
3) The almost complex structure on $Z(M)$ is integrable.
\end{thm}

\begin{example}\label{T}\rm
Let $V$ be a quaternionic vector space of real dimension $4n$. The twistor space of $V$ is $Z(V)\simeq\bC^{2n}\otimes{\mathcal O}_{\bP^1}(1)$ 
\end{example}

Let $M$ be a special K\"ahler manifold and $T^{\ast}M$ its cotangent bundle. From theorem (\ref{sk}) it follows that $T^{\ast}M$ is a quaternionic manifold. Moreover, from lemma (\ref{corre}) we know that on a given vector space $V$, there is a one to one correspondence between Hodge structures of weight one and quaternionic structures.
The relative version of that correspondence reads as follows: let ${\mathcal S}(T^{\bC} M)$  be the family over $M$ whose fibre at 
the point $x$ is the Rees-Simpson bundle associated to the quaternionic space $T_x^{\bC} M$.
\begin{thm}
The pull-back to $T^{\ast}M$ of the family ${\mathcal S}(T^{\bC} M)$  gives the family of the normal bundles to the twistor lines of the $Z(T^\ast M)$.
\end{thm}

Finally, we have the following:

\begin{corollary}\label{mainth}
 The variation of polarized Hodge structures of weight 1 defined on $T^{\bC}M$ induces a quaternionic structure on $T^{\ast}M$. This quaternionic structure coincides with that one of theorem (\ref{sk}).
\end{corollary}

\subsection{Acknowledgments}
The authors express their gratitude to Vincente Cort\'es for useful comments and remarks on a previous version of the paper.

\end{document}